\documentclass[useAMS,usenatbib, twocolumn]{aastex631}

\usepackage{natbib}
\usepackage{graphicx}
\usepackage{epsfig}
\usepackage{amssymb}
\usepackage{amsmath}
\usepackage{natbib}
\usepackage{times}

\usepackage[english]{babel}

\begin{document}
 
\title{Does the Streaming Instability exist within the Terminal Velocity Approximation?}
\shorttitle{Refinement of the terminal velocity approximation}
\author{V. V. Zhuravlev}
\shortauthors{Zhuravlev}
\affiliation{Sternberg Astronomical Institute, Lomonosov Moscow State University, Universitetskij pr., 13, Moscow 119234, Russia}

\defcitealias{zhuravlev_2019}{Z19}
\defcitealias{pan_2021}{P21}

\begin{abstract}
Terminal velocity approximation is appropriate to study the dynamics of gas-dust mixture with solids
tightly coupled to the gas. This work reconsiders its compatibility with physical processes giving 
rise to the resonant Streaming Instability in the low dust density limit. 
It is shown that the linearised equations have been commonly used to study 
the Streaming Instability within the terminal velocity approximation actually 
exceed the accuracy of this approximation. 
For that reason, the corresponding dispersion equation recovers the long wavelength branch 
of the resonant Streaming Instability caused by the stationary azimuthal drift of the dust. 
However, the latter must remain beyond the terminal velocity approximation by its 
physical definition. The refined equations for gas-dust dynamics in the terminal 
velocity approximation does not lead to the resonant Streaming Instability. 
The work additionally elucidates the physical processes responsible for the instability.

\end{abstract}


\section{Introduction}

Solid particles accumulated in the midplane of protoplanetary disk should drift rapidly
towards the host star. Immersed in a rotating gas, this flow serves a source of free energy 
for the local growth of dust overdensities, which may further 
become the sites for planetesimal formation. Study of the corresponding linear instability 
was initiated by \citet{youdin-goodman-2005} who realized that mutual coupling 
of solids with gas by the drag force may lead to amplification of small gas-dust 
perturbations, a process generally similar to the two-stream plasma instability known since 
the pioneering works of \citet{haeff-1948} and \citet{pierce-1948}. By that analogy, 
\citet{youdin-goodman-2005} termed the new gas-dust instability as the Streaming Instability 
(hereafter SI). They showed that the free parameters determining SI are the stopping
time of the particles expressed in units of the Keplerian time (the Stokes number) 
and the dust-to-gas density ratio. The numerical analysis of SI 
has revealed a rather complex pattern of its growth rate. Further analytical developments, 
including the most recent ones, have identified four branches of SI. 

At first, \citet{youdin-goodman-2005} and \citet{latter-2011} 
found that the secular mode associated with the advection of the dust density perturbations 
becomes growing with the account of the dust backreaction on the gas, which 
enhances the pressure perturbation maxima serving as the dust traps. 
These works employed the so called terminal velocity approximation (hereafter TVA). 
TVA naturally arises while considering the gas-dust dynamics in terms of the center-of-mass
velocity of gas-dust mixture and the relative velocity between the dust and the gas.
According to \citet{youdin-goodman-2005}, TVA assumes the marginal balance between
the drag force and the gas pressure force, which both enter the equation 
for the relative velocity of mixture. Additionally, TVA neglects small contribution 
of the relative velocity to equation for the center-of-mass velocity of mixture. 
TVA is usefull for the dynamics of a well-coupled mixture with small Stokes number. 
\citet{laibe-price-2014} generalized the formulation of gas-dust dynamics 
in terms of center-of-mass velocity and relative velocity of the mixture and showed that 
it is optimal for use in numerical simulations. 
This is especially so for the well-coupled mixture, when 
the relative motion of two fluids is described within TVA and mathematically determined 
by a diffusion-like term in the equation for the dust-to-gas ratio evolution. 
This approach made it possible to consider the dusty gas 
with partially coupled solids as the single fluid with heating and cooling, and further, 
the growth of secular mode of SI as its overstability, see \citet{lin-youdin-2017}. 
The one-fluid model, as formulated within TVA, has been employed in numerical studies of gas-dust
dynamics with small solids embedded in protoplanetary disks \citep{lin-2019, lovascio-2019}, 
as well as to consider some extensions of SI, see \citep{chen-lin-2020, paardekooper-2020, lin-2021}.

Another branch of SI is represented by the 
inertial wave\footnote{nicknamed as epicycle mode} of the gas, 
which also becomes growing with the account of the dust backreaction on gas. 
For that, one should retain the higher order terms over the Stokes number, which are beyond TVA. 
This was shown by \citet{jaupart_2020}, 
who additionally found that the terms standing beyond TVA surprisingly contribute to the growth rate 
of the secular mode previously considered only within TVA, 
see also the recent findings of \citet{pan_2021} (hereafter \citetalias{pan_2021}) 
on this issue.

The regime of strong non-linear clumping triggered by SI has been revealed for the first time 
in an unstratified local simulations of \citet{johansen-2007-apj-2}. It was shown that for solids
corresponding to the Stokes number less than unity, which better meets the fragmentation barrier, 
the strong clumping occurs only for the dust-to-gas density ratio significantly higher than unity. 
This feature of the non-linear SI has manifested itself in a more realistic stratified models, 
which included the dust settling, see e.g. \citet{johansen-2009}, \citet{bai-stone-2010},
\citet{carrera-2015}, \citet{johansen-2017}, \citet{li-youdin-2021} and \citet{lesur-rev-2022}
for more detailed review. The analytical study of the branch of the linear SI
associated with the high dust-to-gas density ratio started from \citet{squire_2018} and 
was continued by \citet{squire_2020} and \citet{pan_2020}. It was shown that, contrary to 
the previously known analytic branches of SI, this one is provided by standing perturbations with 
the growth rate independent of the Stokes number.


Along with that, the analytic breakthrough occurred in the opposite limit of low dust fraction.
\citet{squire_2018} identified the fastest-growing modes of SI
on the scales where the velocity of dust radial drift matches the phase velocity of inertial waves
propagating in gas. This led to understanding that in the limit of low dust fraction, 
the most prominent SI is provided by a linear resonance
between the inertial wave and the dust wave, which is represented by the advection of the dust density
perturbations, see \citet{zhuravlev_2019} (hereafter \citetalias{zhuravlev_2019}). 
At the same time, it became clear that the resonant mechanism leading to instability must be subtle.
In the case of low dust fraction, \citet{squire_2020} found that the force caused by the radial drift 
of the dust density perturbations is strictly out of phase with oscillations of gas 
as considered to leading order in the small Stokes number. The adjustment of the distribution of dust 
necessary to recover the instability comes from gas-dust dynamics of the next order in the Stokes number. 
This result appeared to be in accordance with the analysis of \citetalias{zhuravlev_2019},
who showed that the mode coupling giving rise to the resonant drag instability associated with 
the settling of dust is absent when dust drifts only radially in a disk. Since \citetalias{zhuravlev_2019} 
studied the mode coupling in the framework of TVA, 
he concluded that SI must be provided by some other resonant mechanism related to the inertia of
solids, which goes beyond TVA. However, \citetalias{pan_2021} claimed that the general 
dispersion equation constructed from the commonly known linearised two-fluid equations also derived within TVA
do recover the resonant SI in the limit of low dust fraction as well as the small vertical wavenumbers. 
The latter is in evident contradiction with conclusions of \citetalias{zhuravlev_2019}.

The purpose of this study is to resolve a discrepancy between \citetalias{zhuravlev_2019} and
\citetalias{pan_2021} in order to avoid possible further misunderstanding of the physical meaning of TVA.
I argue that claim of \citetalias{pan_2021} is formal because it has no physical grounds.
The limit of the small vertical wavenumbers considered by  \citetalias{pan_2021} 
singles out the long wavelength branch of the resonant SI. However, this branch of the resonant SI is caused by the slow stationary azimuthal drift of the dust, which itself is inertial effect to be neglected in TVA. In order to demonstrate this, 
I revisit the reduced dispersion equation valid in the vicinity of resonance between
the inertial wave and the dust wave in the limit of low dust fraction, see Section 4.2 of
\citetalias{zhuravlev_2019}. Its derivation is intentionally performed using two different sets 
of state variables, in each case, at first within TVA and after that beyond TVA. 
I closely follow the lines of reasoning employed in
\citetalias{zhuravlev_2019} while considering the resonant gas-dust dynamics leading to SI. 
The main conclusion of this study is drawn comparing the four cases mentioned above with each other. 
Finally, I suggest the refined set of general two-fluid equations, 
which should be used in order to study gas-dust dynamics in TVA. 
These equations are used to perform an additional analysis of the general dispersion equation obtained for the arbitrary value of the dust fraction. This analysis is relegated to the
Appendix. It is shown that there is no more growing secular mode in the revised TVA. 
In contrast, the dust-rich branch of SI, which is growing for dust-to-gas ratio larger than
unity, persists in the revised TVA.

In Sections \ref{sec_resonance}-\ref{sec_mixture} and in the Appendix
the gas-dust flow is assumed to be axisymmetric.
Throughout this text I use the original definitions and notations of \citetalias{zhuravlev_2019}.

\section{Resonance between modes}

\label{sec_resonance}

Following \citetalias{zhuravlev_2019}, I consider gas-dust dynamics in the vicinity of 
resonance between the streaming dust wave (SDW) and the inertial wave (IW)\footnote{Another representative example
of such an analysis can be found in \citet{zhuravlev_2021}.}.
IW is a wave that propagates through the interior of rotating gas due to the presence of Coriolis force, see e.g. \citet{landau-lifshitz-1987}. SDW is a trivial mode produced by the stationary drift of the 
dust transporting perturbations of dust density.

In this case, perturbations taken in the form of monochromatic waves 
$\propto \exp (-{\rm i}\omega t + {\rm i}k_x x + {\rm i} k_z z )$
obey the following dispersion equation 
\begin{equation}
\label{disp}
D_g(\omega,{\bf k}) \cdot D_p(\omega,{\bf k}) = \epsilon_m(\omega, {\bf k}),
\end{equation}
where 
\begin{equation}
\label{D_g}
D_g(\omega,{\bf k}) \equiv \omega^2 - \omega_i^2,
\end{equation}
independently describes IW, while 
\begin{equation}
\label{D_p}
D_p(\omega,{\bf k}) \equiv \omega - \omega_p
\end{equation}
independently describes SDW with definitions 
$$
\omega_p \equiv - g_x t_s k_x,
$$
$$
\omega_i \equiv \kappa \frac{k_z}{k},
$$
with $k^2 \equiv k_x^2 + k_z^2$. 
The variables $\kappa$, $t_s$ and $g_x$ introduce, respectively,
the local epicyclic frequency in a disk, the stopping time of the 
particles and the absolute value of effective gravity along the radial direction, 
see \citetalias{zhuravlev_2019} for the details.

Consideration of the dynamical dust back-reaction on gas leads to the appearance of non-zero coupling term in
the right-hand side (RHS) of equation (\ref{disp}). 
As far as the dust fraction $f\ll 1$, the coupling term is small, which means that 
significant change of solution of the dispersion relation occurs only close to the resonance between SDW and IW. 
Below, the subscript after $\epsilon$ introduces a sequence number of model considered in this work. 
At the same time, it can be checked that for all models the left-hand side (LHS) of equation (\ref{disp}), 
which can be taken to the zeroth order in $f$ near the resonance, remains the same. 

The resonance between SDW and IW is determined by the condition of the mode crossing,
\begin{equation}
\label{res_cond}
\omega_p = -\omega_i \equiv \omega_c,
\end{equation}
and it is assumed for simplicity that $k_{x,z}>0$ hereafter. All other cases can be considered in a similar way.

According to equation (\ref{res_cond}), the length-scale of mode crossing is determined by the relation
\begin{equation}
\label{res_scale}
\frac{k_x}{k_z} = \frac{\tilde k}{k}, 
\end{equation}
where 
\begin{equation}
\label{tilde_k}
\tilde k \equiv \frac{\kappa}{g_x t_s}
\end{equation}
is the characteristic wavenumber of resonance between SDW and IW and so the typical 
wavenumber associated with resonant SI.

The corresponding variation of the frequencies of SDW and IW at the mode crossing 
is estimated in the following way,
see the details in \citetalias{zhuravlev_2019},
\begin{equation}
\label{Delta}
\Delta_m = \pm \left ( \frac{\epsilon_m |_{\omega_c}}{\partial_\omega D_g|_{\omega_c} \cdot \partial_\omega D_p|_{\omega_c}} \right )^{1/2} = 
\pm \left ( \frac{\epsilon_m |_{\omega_c}}{2\omega_c} \right)^{1/2}.
\end{equation}
As soon as $\Delta$ has positive imaginary part, the corresponding model gives rise to SI.

\section{Problem considered with respect to perturbations of gas velocity}

\label{sec_gas}

\subsection{General equations}

Local dynamics of gas-dust mixture in a disk on the scales smaller than the disk scale-height 
can be described in the shearing sheet approximation, see \citet{squire_2018} and 
\citetalias{zhuravlev_2019}.
Equation for velocity of gas including dust aerodynamical back-reaction on gas reads
\begin{equation}
\label{eq_U_g}
\begin{aligned}
\partial_t {\bf U}_g - 2\Omega_0 U_{g,y} {\bf e}_x + \frac{\kappa^2}{2\Omega_0} U_{g,x} {\bf e}_y + \\
({\bf U}_g \nabla){\bf U}_g = - \frac{\nabla p}{\rho_g} + \frac{\rho_p}{\rho_g} \frac{\bf V}{t_s},
\end{aligned}
\end{equation}
and it is assumed that the gas density, $\rho_g$, is constant. Hence, gas velocity is strictly free of divergence,
\begin{equation}
\label{eq_rho}
\nabla \cdot {\bf U}_g = 0.
\end{equation}
On the other side, the behavior of dust can be described by the evolution of its density distribution 
using the continuity equation for gas-dust mixture,
\begin{equation}
\label{eq_rho_tot}
\partial_t \rho_p + \nabla ( \rho {\bf U} ) = 0,
\end{equation}
where, by definition,
\begin{equation}
\label{U_g_V}
{\bf U} = {\bf U}_g + \frac{\rho_p}{\rho} {\bf V}
\end{equation}
is the center-of-mass velocity of gas-dust mixture and $\rho\equiv \rho_g+\rho_p$ is the total density of mixture.
The closing equation for the relative velocity of gas and dust, ${\bf V}\equiv {\bf U}_p -{\bf U}_g$, reads
\begin{equation}
\label{eq_V}
\begin{aligned}
\partial_t {\bf V} 
- 2\Omega_0 V_y {\bf e}_x + 
\frac{\kappa^2}{2\Omega_0} V_x {\bf e}_y + \\ ({\bf U}\nabla) {\bf V}  + ({\bf V}\nabla) {\bf U} + 
\frac{\rho_g}{\rho} ({\bf V} \nabla) \left ( \frac{\rho_g}{\rho} {\bf V} \right ) - \\ 
\frac{\rho_p}{\rho} ({\bf V} \nabla) \left ( \frac{\rho_p}{\rho} {\bf V} \right ) 
 =  \frac{\nabla (p+p_0)}{\rho_g} - \frac{\rho}{\rho_g} \frac{{\bf V}}{t_s}.
\end{aligned}
\end{equation}

Equation (\ref{eq_V}) follows from the combination of independent equations for velocities of gas and dust and fully describes the local relative motion of gas and dust in a disk, see Section 2 of \citetalias{zhuravlev_2019}. 

Below in this Section, equations (\ref{eq_U_g}-\ref{eq_rho_tot}, \ref{eq_V}) along with 
the definition (\ref{U_g_V}) are used to derive the set of linear
equations, which represent the dynamics of small gas-dust perturbations in the vicinity of resonance between 
SDW and IW provided that the dust suffers stationary radial drift.

Everywhere below it is assumed that 
\begin{equation}
\label{frac}
f \equiv \frac{\rho_p}{\rho_g} \ll 1
\end{equation}
as well as the Stokes number 
\begin{equation}
\label{tau}
\tau \equiv \Omega_0 t_s \ll 1.
\end{equation}

\vspace{0.2cm}

\subsection{Stationary drift of the dust}

\subsubsection{The solution within TVA}

According to \citet{youdin-goodman-2005}, who first used TVA, ``this approximation, which ignores inertial accelerations'', ``amounts to neglecting all terms on the left-hand side of equation (here \ref{eq_V}), 
both in equilibrium and in perturbation''.
Assuming for simplicity
\begin{equation}
\label{bg_U_g_TVA}
{\bf U}_g = 0,
\end{equation}
one obtains from equations (\ref{eq_U_g}) and (\ref{eq_V}) with their LHS set to zero that,
to leading order in small $f$,
\begin{equation}
\label{bg_V_g_TVA}
{\bf V} = - t_s g_x {\bf e}_x,
\end{equation}
implying that $\rho_p=const$.
Solids drift exactly along the pressure gradient in a disk, which is manifestation of TVA:
``drag forces adjust quasi-statically to pressure forces'' as stated by \citet{youdin-goodman-2005}.

\subsubsection{The solution beyond TVA}

The stationary solution is extended up to the next order in $t_s$ taking into account 
the leading inertial term on LHS of equation (\ref{eq_V}),
which is a weak relative Coriolis acceleration of gas-dust mixture along the azimuthal direction
caused by the radial drift of solids. 
To leading order in small $f$, one obtains
\begin{equation}
\label{bg_U_g}
{\bf U}_g = f t_s g_x {\bf e}_x,
\end{equation}
whereas
\begin{equation}
\label{bg_V_g}
{\bf V} = - t_s g_x {\bf e}_x + \frac{\kappa^2}{2\Omega_0} t_s^2 g_x {\bf e}_y,
\end{equation}
implying again that $\rho_p=const$.
Note that the terms $\sim t_s^3$ an smaller have been omitted in equations (\ref{bg_U_g}-\ref{bg_V_g}). Clearly, weak azimuthal drift of the dust found in the stationary solution 
goes beyond TVA, as soon as the corresponding drag force is adjusted by no pressure force.

\subsection{Linear dynamics of gas-dust perturbations within TVA}

%

The state variables, which determine the behavior of corresponding gas-dust perturbations are 
the Eulerian perturbations of gas velocity, ${\bf u}_g$, enthalpy, 
$W_g \equiv p^\prime / \rho_g$, and the relative density of dust, $\delta \equiv \rho_p^\prime /\rho_p$, 
where $p^\prime$, $\rho_p^\prime$ and $\rho_p$ are, respectively, the Eulerian perturbations of pressure 
and density of dust the background density of dust.

Equations for perturbations on the background (\ref{bg_U_g_TVA}-\ref{bg_V_g_TVA}) valid in the vicinity of 
resonance between SDW and IW read 
\begin{equation}
\label{eq_g_TVA_1}
\partial_t u_{g,x} - 2\Omega_0 u_{g,y} = -\partial_x W_g + f \frac{v_x}{t_s} - f g_x \delta,
\end{equation}

\begin{equation}
\label{eq_g_TVA_2}
\partial_t u_{g,y} + \frac{\kappa^2}{2\Omega_0} u_{g,x} = 0,
\end{equation}

\begin{equation}
\label{eq_g_TVA_3}
\partial_t u_{g,z} = -\partial_z W_g + f \frac{v_z}{t_s},
\end{equation}

\begin{equation}
\label{eq_g_TVA_4}
\partial_x u_{g,x} + \partial_z u_{g,z} = 0,
\end{equation}

\begin{equation}
\label{eq_g_TVA_5}
\partial_t \delta = - \partial_{x} v_x - \partial_{z} v_z + t_s g_x \partial_x \delta,
\end{equation}
where the terms $\sim O(f)$ and smaller entering equation for dust, (\ref{eq_g_TVA_5}), have been omitted.
The Eulerian perturbation of the relative velocity of mixture, ${\bf v}$, is taken from equation
(\ref{eq_V}) employing TVA to the zeroth order in $f$. This is
\begin{equation}
\label{eq_g_TVA_6}
v_x = t_s \partial_x W_g,
\end{equation}
\begin{equation}
\label{eq_g_TVA_7}
v_z = t_s \partial_z W_g.
\end{equation}


Further, it is appropriate to introduce the new variables 
for the incompressible motion of gas:
\begin{equation}
\label{Varpi_g}
\varpi_g \equiv -\partial_z u_{g,y} \nonumber
\end{equation}
and
\begin{equation}
\label{Phi_g}
\phi_g \equiv \partial_z u_{g,x}, \nonumber
\end{equation}
which, along with $u_{g,z}$ and $\delta$, obey 
the concise set of equations:
\begin{equation}
\label{Sys_fin_g_TVA_1}
\partial_t \phi_g = \partial_{tx} u_{g,z}  - 2\Omega_0 \varpi_g - f g_x \partial_z \delta, 
\end{equation}
\begin{equation}
\label{Sys_fin_g_TVA_2}
\partial_t \varpi_g = \frac{\kappa^2}{2\Omega_0} \phi_g,
\end{equation}
\begin{equation}
\label{Sys_fin_g_TVA_3}
\partial_{tx}\varpi_g = - \frac{\kappa^2}{2\Omega_0} \partial_{zz} u_{g,z},
\end{equation}
\begin{equation}
\label{Sys_fin_g_TVA_4}
\partial_{tz}\delta = t_s g_x \partial_{xz} \delta + 2\tau \partial_x \varpi_g. 
\end{equation}

In order to obtain equations (\ref{Sys_fin_g_TVA_1}-\ref{Sys_fin_g_TVA_3}) one takes the curl of equations 
(\ref{eq_g_TVA_1}-\ref{eq_g_TVA_3}). Equation (\ref{Sys_fin_g_TVA_4}) is obtained 
from equations (\ref{eq_g_TVA_5}-\ref{eq_g_TVA_7}) after taking the divergence of equations 
(\ref{eq_g_TVA_1}-\ref{eq_g_TVA_3}) to the zeroth order in $f$.
Note that in equation (\ref{Sys_fin_g_TVA_4}) and elsewhere below the combination $\Omega_0 t_s$ is replaced by the Stokes number according to definition (\ref{tau}).

Equations (\ref{Sys_fin_g_TVA_1}-\ref{Sys_fin_g_TVA_4}) written for modes of perturbations 
yield the dispersion equation in the form given by equation (\ref{disp}). The coupling term
emerges due to the non-zero product of the corresponding terms originating from the last 
term on RHS of equation (\ref{Sys_fin_g_TVA_1}) and the last term on RHS of equation
(\ref{Sys_fin_g_TVA_4}): 
\begin{equation}
\label{coupling_1}
\epsilon_1 = - f t_s \kappa^2 g_x k_x \frac{k_z^2}{k^2}.
\end{equation}
According to equation (\ref{Delta}), the variation of frequency 
at resonance between SDW and IW provided by $\epsilon_1$
is real at $\omega_c = \omega_p = -g_x t_s k_x$.
Explicitly,
\begin{equation}
\label{Delta_1}
\Delta_1 = \pm \kappa\, \left ( \frac{f}{2} \right )^{1/2} \frac{k_z}{k}
\end{equation}
resulting in the {\it absence of instability}.

\subsection{Linear dynamics of gas-dust perturbations beyond TVA}
\label{sec_g}

In this case, equations (\ref{eq_g_TVA_1}-\ref{eq_g_TVA_4}) for gas change only due to modification 
of the relative velocity of mixture in the stationary solution. Indeed, revision of LHS of equation 
(\ref{eq_V}) shows that it cannot give any contributions $\sim \delta$ to the zeroth order in $f$, which
could enter RHS of equations (\ref{eq_g_TVA_1}-\ref{eq_g_TVA_3}) via ${\bf v}$ over there. 
Hence, equation (\ref{bg_V_g}) yields

\begin{equation}
\label{eq_g_2}
\partial_t u_{g,y} + \frac{\kappa^2}{2\Omega_0} u_{g,x} = f \frac{\kappa^2}{2\Omega_0} t_s g_x \delta,
\end{equation}
instead of equation (\ref{eq_g_TVA_2}). The new term on RHS of this equation comes from weak azimuthal drift 
of the dust.

%
%

The divergence of ${\bf v}$ on RHS of equation (\ref{eq_g_TVA_5}) should be expressed 
in terms of the gas variables.
For that, equations (\ref{eq_g_TVA_6}) and (\ref{eq_g_TVA_7}) are replaced by an augmented equations
\begin{equation}
\label{eq_g_6}
-2\Omega_0 v_y + ({\bf V \nabla}) u_{g,x} = \partial_x W_g - \frac{v_x}{t_s}
\end{equation}
and
\begin{equation}
\label{eq_g_7}
({\bf V \nabla}) u_{g,z} = \partial_z W_g - \frac{v_z}{t_s},
\end{equation}
respectively, check LHS of equation (\ref{eq_V}) to the zeroth order in $f$.
Note that the terms $\partial_t {\bf v}$ and $({\bf V}\nabla) {\bf v}$ on LHS of equation (\ref{eq_V})
written for perturbations cancel each other near the resonance up to the small difference $\sim O(f)$, 
which is omitted here. The replacement ${\bf u} \to {\bf u}_{g}$ 
in the advection terms on LHS of equations (\ref{eq_g_6}-\ref{eq_g_7}) has been done by the same reason.

Only the first term on LHS of equation (\ref{eq_g_6}) contributes to $\nabla \cdot {\bf v}$ 
and, thus, to contraction/rarefaction of dust.
This term represents perturbation of the relative Coriolis acceleration of a mixture 
due to the non-zero perturbation of the azimuthal relative velocity of mixture.
The emergence of the azimuthal relative velocity of mixture 
makes one to consider the dynamics of a mixture in azimuthal direction.

It is important to note that within TVA the azimuthal motion of dust in the axisymmetric 
perturbed flow must be trivial: as soon as the pressure gradient has no projection 
onto the azimuthal direction, inertia-free solids perfectly follow azimuthal motion of the gas, 
and $v_y=0$.
However, the synchronicity of gas and dust azimuthal motions is broken beyond TVA 
due to the azimuthal balance of forces coming in the next-order over $t_s$ 
(but still considered to the zeroth order in $f$), see equation (\ref{eq_V}) :
\begin{equation}
\label{eq_g_6p}
t_s g_x \partial_x u_{g,y} = \frac{v_y}{t_s}.
\end{equation}
Physically\footnote{It can be checked that the 'Coriolis' term, $\kappa^2/(2\Omega_0) v_x {\bf e}_y$, 
standing also on LHS of equation (\ref{eq_g_6p}) in the zeroth order in $f$ 
is small compared with the term $(\bf V \nabla) u_{g,y}$ over there.}, equation (\ref{eq_g_6p}) describes the
azimuthal acceleration of solids by the gas drag as solids drift through the perturbed shear flow
of the gas.
This is purely inertial effect, since the inertia-free particles would be instantly 
lifted up by the gas producing no relative velocity. 

Equations (\ref{eq_g_6}-\ref{eq_g_7}) yield
\begin{equation}
\label{div_v}
\nabla \cdot {\bf v} = t_s \nabla^2 W_g + 2\tau \partial_x v_y,
\end{equation}
where $\nabla^2 \equiv \partial_{xx}+\partial_{zz}$. Thus, an additional compression/rarefaction of dust  
beyond TVA originates from the {\it relative azimuthal motion between gas and dust}, which in turn, 
emerges due to the azimuthal acceleration of gas as seen for solids penetrating the gas eddies in the course of 
their bulk radial drift (cf. Section 4 of \citet{squire_2020}). Equation (\ref{div_v}) in combination 
with equation (\ref{eq_g_6p}) is used to obtain an augmented version of equation (\ref{eq_g_TVA_5}). 


The final equations read
\begin{equation}
\label{Sys_fin_g_1}
\partial_t \phi_g = \partial_{tx} u_{g,z}  - 2\Omega_0 \varpi_g - f g_x \partial_z \delta, 
\end{equation}
\begin{equation}
\label{Sys_fin_g_2}
\partial_t \varpi_g = \frac{\kappa^2}{2\Omega_0} \phi_g - f t_s g_x \frac{\kappa^2}{2\Omega_0} \partial_z \delta,
\end{equation}
\begin{equation}
\label{Sys_fin_g_3}
\partial_{tx}\varpi_g = - \frac{\kappa^2}{2\Omega_0} \partial_{zz} u_{g,z} - f t_s g_x \frac{\kappa^2}{2\Omega_0} \partial_{xz} \delta,
\end{equation}
\begin{equation}
\label{Sys_fin_g_4}
\partial_{tz}\delta = t_s g_x \partial_{xz} \delta + 2\tau \partial_x \varpi_g + 2\tau g_x t_s^2 \partial_{xx} \varpi_g. 
\end{equation}

Equations (\ref{Sys_fin_g_1}-\ref{Sys_fin_g_4}) yield the generalized coupling term
\begin{equation}
\label{coupling_2}
\epsilon_2(\omega, {\bf k}) \equiv - f t_s \kappa^2 g_x k_x \frac{k_z^2}{k^2} 
\left ( 1 + {\rm i} g_x k_x t_s^2 - {\rm i} \omega t_s \frac{k^2}{k_z^2}  \right )
\end{equation}
and the corresponding new correction to the frequency of mode crossing
\begin{equation}
\label{Delta_2}
\Delta_2 \approx 
\pm \kappa \left ( \frac{f}{2} \right )^{1/2} \frac{k_z}{k} \left [ 1 + 
{\rm i}\, \frac{\kappa t_s}{2} \frac{k_z}{k} \left ( 1 + \frac{k^2}{k_z^2} \right ) \right ],
\end{equation}
which fully recovers SI, see equation (5.10) of \citet{squire_2018} as well as equation (109) of 
\citetalias{zhuravlev_2019}.

It is important to note that the first and the last terms after the imaginary unit come from the non-zero 
LHS of equation (\ref{eq_g_6p}) and RHS of equation (\ref{eq_g_2}), respectively. 
This means that generally resonant SI is provided by a combination of two effects:

i) the spin up of the gas eddies by the stationary radial drift of solids, 
undergoing a weak spatial redistribution due 
to their perturbed azimuthal motion relative to the gas as solids penetrate 
the gas eddies in the radial direction. \\
ii) the spin up of the gas eddies by a weak stationary azimuthal drift of solids; 
the subsequent growth of pressure gradient enhances the accumulation of solids, which further enhances an
azimuthal spin up of eddies leading to instability;
\\

Both of these effects are inertial in the sense that they become stronger with the increasing mass of the particles, i.e. their stopping time (cf. Section 5 of \citetalias{zhuravlev_2019}).

\subsubsection{The long wavelength limit}
\label{lw_limit}

The long wavelength limit of resonant SI corresponds to the limit of small $k$ as compared with the 
characteristic {wavenumber} of SI, $\tilde k$, see equations (\ref{res_scale}) and (\ref{tilde_k}).
As soon as $k \ll \tilde k$, the resonant waves propagate almost radially, $k_x \gg k_z$ and $k_x \approx k$.
In this case, the leading term producing SI in equation (\ref{Delta_2}) is the last one after the imaginary unit.
Accordingly, one obtains
\begin{equation}
\label{long_wave_Delta}
\Delta_2 \to \pm \left ( \frac{f}{2} \right )^{1/2} \left [ \frac{k_z}{k} + {\rm i} \frac{\kappa t_s}{2} \right ],
\end{equation}
which recovers the result of \citetalias{pan_2021}, see his equation (29) taken in the same limit, $k_x \gg k_z$. Derivation of equation (\ref{Delta_2}) demonstrates that this regime of the resonant 
SI is associated with the stationary azimuthal drift of solids. 
However, the stationary azimuthal drift of solids goes beyond TVA because it is caused by the 
weak difference of Coriolis force acting on solids and gas, rather than by the pressure gradient.
In turn, this implies that equations (22-26) from \citet{latter-2011} commonly known as equations derived within TVA actually exceed the accuracy of TVA. The next Section addresses this issue.


\section{Problem considered with respect to perturbations of gas-dust mixture velocity}

\label{sec_mixture}

\subsection{General equations}

Local dynamics of gas-dust mixture can be equivalently described using equation for the center-of-mass
velocity of mixture rather than equation for the velocity of gas.

This is taken from \citetalias{zhuravlev_2019}:
\begin{equation}
\label{eq_U}
\begin{aligned}
\partial_t {\bf U} - 2\Omega_0 U_y {\bf e}_x + \frac{\kappa^2}{2\Omega_0} U_x {\bf e}_y + ({\bf U}\nabla) {\bf U} + \\ 
\frac{\rho_g}{\rho} \left \{  \left ( {\bf V} \nabla \left ( \frac{\rho_p}{\rho} \right ) \right ) {\bf V} +
2\frac{\rho_p}{\rho} \left ( {\bf V} \nabla \right ) {\bf V} \right \} =\\ 
\frac{\nabla p_0}{\rho_g} - \frac{\nabla(p+p_0)}{\rho}.
\end{aligned}
\end{equation}
The key point is that ${\bf U}$ is not free of divergence but
\begin{equation}
\label{eq_rho_g}
\nabla \cdot {\bf U} = \nabla \cdot \left ( \frac{\rho_p}{\rho}{\bf V} \right ).
\end{equation}

As previously, evolution of the dust density and the relative velocity of mixture are governed 
by the equations (\ref{eq_rho_tot}) and (\ref{eq_V}), respectively.

Below in this Section, 
equations (\ref{eq_U}-\ref{eq_rho_g}) along with (\ref{eq_rho_tot}) and (\ref{eq_V}) 
are used to reproduce the derivation of linear equations for dynamics of small gas-dust perturbations 
in the vicinity of resonance between SDW and IW in the presence of stationary radial drift of the dust.

\subsection{Stationary drift of the dust}

\subsubsection{The solution within TVA}

Assuming for simplicity that

\begin{equation}
\label{bg_U_TVA}
{\bf U} = 0,
\end{equation}
one obtains from equations (\ref{eq_U}) and (\ref{eq_V}) that 
\begin{equation}
\label{bg_V_TVA}
{\bf V} = - t_s g_x {\bf e}_x,
\end{equation}
implying that $\rho_p=const$. Note that equations 
(\ref{bg_V_TVA}) and (\ref{bg_V_g_TVA}) are identical to each other.
The solution (\ref{bg_U_TVA}-\ref{bg_V_TVA}) has been used by \citetalias{zhuravlev_2019}, 
see his equations (22-24).

\subsubsection{The solution beyond TVA}

Assuming again that
\begin{equation}
\label{bg_U}
{\bf U} = 0,
\end{equation}
one obtains to leading order in small $f$
\begin{equation}
\label{bg_V}
{\bf V} = - t_s g_x {\bf e}_x + \frac{\kappa^2}{2\Omega_0} t_s^2 g_x {\bf e}_y,
\end{equation}
implying again that $\rho_p=const$.
The terms $\sim t_s^3$ an smaller have been omitted in equation (\ref{bg_V}).
Note that equations (\ref{bg_V}) and (\ref{bg_V_g}) are identical to each other.
The solution (\ref{bg_U}-\ref{bg_V}) has been used 
by \citetalias{zhuravlev_2019} in order to consider SI, 
see his equations (86-89)\footnote{Note that equation (89) of \citetalias{zhuravlev_2019} contains a misprint}.

\subsection{Linear dynamics of gas-dust perturbations within TVA}

In this Section, I replace equations (\ref{eq_g_TVA_1}-\ref{eq_g_TVA_4}) by the equations (\ref{eq_U}) and 
(\ref{eq_rho_g}) linearised on the background (\ref{bg_U_TVA}-\ref{bg_V_TVA}):

\begin{equation}
\label{eq_1_TVA}
\partial_t u_{x} - 2\Omega_0 u_{y} = -\partial_x W - f g_x \delta,
\end{equation}

\begin{equation}
\label{eq_2_TVA}
\partial_t u_{y} + \frac{\kappa^2}{2\Omega_0} u_{x} = 0,
\end{equation}

\begin{equation}
\label{eq_3_TVA}
\partial_t u_{z} = -\partial_z W,
\end{equation}

\begin{equation}
\label{eq_4_TVA}
\partial_x u_{x} + \partial_z u_{z} = f (\partial_{x} v_x + \partial_{z} v_z) - f t_s g_x \partial_x \delta,
\end{equation}
where ${\bf u}$ and $W \equiv p^\prime / \rho$ are, respectively, the Eulerian perturbation of the center-of-mass
velocity and the Eulerian perturbation of generalized enthalpy. 
Equations (\ref{eq_1_TVA}-\ref{eq_4_TVA}) are
supplemented by an equations describing the dust, which are identical to equations 
(\ref{eq_g_TVA_5}-\ref{eq_g_TVA_7}) with the replacement $W_g \to W$, what can be safely done to the 
zeroth order in $f$.

Introducing the appropriate variables 
\begin{equation}
\label{Varpi}
\varpi \equiv -\partial_z u_{y}, \nonumber
\end{equation}
and
\begin{equation}
\label{Phi}
\phi \equiv \partial_z u_{x} \nonumber
\end{equation}
one arrives at equations
\begin{equation}
\label{Sys_fin_TVA_1}
\partial_t \phi = \partial_{tx} u_{z}  - 2\Omega_0 \varpi - f g_x \partial_z \delta, 
\end{equation}
\begin{equation}
\label{Sys_fin_TVA_2}
\partial_t \varpi = \frac{\kappa^2}{2\Omega_0} \phi,
\end{equation}
\begin{equation}
\label{Sys_fin_TVA_3}
\partial_{tx}\varpi = - \frac{\kappa^2}{2\Omega_0} \partial_{zz} u_{z} - f t_s g_x \frac{\kappa^2}{2\Omega_0} \partial_{xz} \delta,
\end{equation}
\begin{equation}
\label{Sys_fin_TVA_4}
\partial_{tz}\delta = t_s g_x \partial_{xz} \delta + 2\tau \partial_x \varpi. 
\end{equation}
Note that in order to derive equation (\ref{Sys_fin_TVA_3}), the term $\sim \nabla \cdot {\bf v}$, which also
contributes to $\nabla \cdot {\bf u}$ via equation (\ref{eq_4_TVA}), has been omitted since
it does not contribute any term $\sim f\delta$ into equation (\ref{Sys_fin_TVA_3}).

It is not difficult to see that 
equations (\ref{Sys_fin_TVA_1}-\ref{Sys_fin_TVA_4}) are {\it not} equivalent to 
equations (\ref{Sys_fin_g_TVA_1}-\ref{Sys_fin_g_TVA_4}) neither with an account for the replacement 
${\bf u} \leftrightarrow {\bf u}_g$, nor with an account for small difference $\sim O(f)$ 
between ${\bf U}$ and ${\bf U}_g$ in the stationary solutions\footnote{In the latter case, this leads to 
new terms $\sim O(f)$ in $D_g$ and $D_p$ which do not contribute to the resonant solution}.

Equations (\ref{Sys_fin_TVA_1}-\ref{Sys_fin_TVA_4}) yield the new variant of the coupling term,
\begin{equation}
\label{coupling_3}
\epsilon_3(\omega, {\bf k}) \equiv - f t_s \kappa^2 g_x k_x \frac{k_z^2}{k^2} 
\left ( 1 - {\rm i} \omega t_s \frac{k_x^2}{k_z^2}  \right ),
\end{equation}
and the corresponding variant of correction to the frequency of mode crossing
\begin{equation}
\label{Delta_3}
\Delta_3 \approx 
\pm \kappa \left ( \frac{f}{2} \right )^{1/2} \frac{k_z}{k} \left [ 1 + 
{\rm i} \frac{\kappa t_s}{2} \frac{k_x^2}{k k_z}  \right ],
\end{equation}
which fully recovers equation (29) of \citetalias{pan_2021} and indeed recovers the long wavelength limit 
of the resonant SI growth rate, see equation (\ref{long_wave_Delta}).
It can be seen that the growth rate found in equation (\ref{Delta_3}) arises while retaining the 
last term on RHS of equation (\ref{Sys_fin_TVA_3}). In the absence of this term equations 
(\ref{Sys_fin_TVA_1}-\ref{Sys_fin_TVA_4}) become identical to equations 
(\ref{Sys_fin_g_TVA_1}-\ref{Sys_fin_g_TVA_4}) after the replacement ${\bf u} \to {\bf u}_g$ and produce 
no growing modes. This is equivalent to assumption that 
\begin{equation}
\label{div_u_0}
\nabla \cdot {\bf u} = 0
\end{equation}
as supplementary condition for equations (\ref{eq_1_TVA}-\ref{eq_3_TVA}). 
However, ${\bf u}$ cannot be considered free of divergence in order to derive equation for evolution 
of the dust density starting from the general equation (\ref{eq_rho_tot}).
The assumption (\ref{div_u_0}) has been applied in order to derive equations 
(26-29) in \citetalias{zhuravlev_2019}, however, this is not the case for formal solution of equations 
(22-26) from \citet{latter-2011}. 
Section \ref{sec_gen_eq} below is reserved for formulation of the refined 
general two-fluid equations for gas-dust dynamics within TVA.

\subsection{Linear dynamics of gas-dust perturbations beyond TVA}

For completeness, this Section demonstrates how the full growth rate of resonant SI already 
recovered in Section \ref{sec_g} is reproduced using alternative variables.

As was shown in Section 4.2 of \citetalias{zhuravlev_2019}, 
new terms $\sim f\delta$ appearing in equation for ${\bf u}$ to the higher order in $t_s$ originate 
from the first gradient term $\sim V^2$ on LHS of equation (\ref{eq_U}).
Contrary to what occurs with equation for ${\bf u}_g$, see Section \ref{sec_g}, an extension of the stationary 
solution beyond TVA does not bring any terms in equation for ${\bf u}$. 
As a result, equation (\ref{eq_1_TVA}) should be replaced by the following one:
\begin{equation}
\label{eq_1}
\partial_t u_{x} - 2\Omega_0 u_{y} + f t_s^2 g_x^2 \partial_x \delta = -\partial_x W - f g_x \delta,
\end{equation}

On the other side, the dynamics of dust is again described by an equation (\ref{eq_g_TVA_5}).
The divergence of ${\bf v}$ can be excluded from equation (\ref{eq_g_TVA_5}) 
proceeding along the lines of Section \ref{sec_g}, after the replacements $W_g\to W$, ${\bf u}_{g} \to {\bf u}$
done back in equations (\ref{eq_g_6}), (\ref{eq_g_7}) and (\ref{eq_g_6p}) valid to the zeroth order in $f$.

The new set of equations in the appropriate variables reads
\begin{equation}
\label{Sys_fin_1}
\partial_t \phi = \partial_{tx} u_{z}  - 2\Omega_0 \varpi - f g_x \partial_z \delta 
- f t_s^2 g_x^2 \partial_{xz} \delta, 
\end{equation}
\begin{equation}
\label{Sys_fin_2}
\partial_t \varpi = \frac{\kappa^2}{2\Omega_0} \phi,
\end{equation}
\begin{equation}
\label{Sys_fin_3}
\partial_{tx}\varpi = - \frac{\kappa^2}{2\Omega_0} \partial_{zz} u_{z} - f t_s g_x \frac{\kappa^2}{2\Omega_0} \partial_{xz} \delta,
\end{equation}
\begin{equation}
\label{Sys_fin_4}
\partial_{tz}\delta = t_s g_x \partial_{xz} \delta + 2\tau \partial_x \varpi + 2\tau g_x t_s^2 \partial_{xx} \varpi. 
\end{equation}

Note that equations (\ref{Sys_fin_1}-\ref{Sys_fin_4}) differ from equations (\ref{Sys_fin_g_1}-\ref{Sys_fin_g_4}).
Yet, they yield the following coupling term
\begin{equation}
\label{coupling_4}
\epsilon_4(\omega, {\bf k}) \equiv - f t_s \kappa^2 g_x k_x \frac{k_z^2}{k^2} 
\left ( 1 + 2 {\rm i} g_x k_x t_s^2 - {\rm i} \omega t_s \frac{k_x^2}{k_z^2}  \right ),
\end{equation}
which provides dispersion equation with correction to the frequency of mode
crossing {\it identical} to equation (\ref{Delta_2}),
\begin{equation}
\label{Delta_4}
\Delta_4 = \Delta_2.
\end{equation}

\section{Refinement of TVA}
\label{sec_gen_eq}

Additional justification of the condition (\ref{div_u_0}) 
can be found from the ordering of terms in the set of general equations 
(\ref{eq_U}-\ref{eq_rho_g}), (\ref{eq_rho_tot}) and (\ref{eq_V}). 
Such a procedure was carried out in Section 2.1 of
\citetalias{zhuravlev_2019}. It was concluded that TVA corresponds to the limit when all 
terms of orders of $\tau_*$, $\lambda^{-1}$ and $\sqrt{\lambda^{-1}}$ standing in equations 
(\ref{eq_U}) and (\ref{eq_V}) should be omitted. Here 
\begin{equation}
\label{TVA_cond_1}
\tau_* \equiv \max\{ t_s t_{ev}^{-1},\, \tau \} \ll 1
\end{equation}
and
\begin{equation}
\label{TVA_cond_2}
\lambda^{-1} \equiv \frac{g t_s^2}{l_{ev}} \ll 1
\end{equation}
are small dimensionless parameters with $g$, $l_{ev}$ and $t_{ev}$ being, respectively, 
the characteristic specific pressure gradient, length-scale and time-scale of gas-dust mixture dynamics. Note that the characteristic length-scale associated with resonant SI, $\tilde k^{-1}$, corresponds to $\lambda^{-1} \sim \tau \ll 1$ in Keplerian disks 
as follows from equation (\ref{tilde_k}).

However, the use of equation (\ref{eq_U}) taken in TVA, in combination with equation (\ref{eq_rho_g})
{\it leads to an excess of accuracy}. Indeed, as soon as the characteristic absolute value of ${\bf V}$ 
is by factor of $\sqrt{\lambda^{-1}}$ smaller than that of ${\bf U}$, see \citetalias{zhuravlev_2019}, 
the inertial terms on LHS of equation (\ref{eq_U}) rearranged with the help of equation (\ref{eq_rho_g})
also retain an extra term of order of $\tau_*$. An example of such an extra term is found in
equation (\ref{Sys_fin_TVA_3}), which is the last one on its RHS. 
It is also clear why this term makes it possible to recover the correct resonant SI growth rate 
in the long wavelength 
limit. The condition $k\ll \tilde k$ implies that 
\begin{equation}
\label{long_wl}
\lambda^{-1} \ll \tau
\end{equation}
according to definitions of 
$\tilde k$ and $\lambda$. So, a small term on RHS of equation (\ref{Sys_fin_TVA_3}) prevails upon 
perturbations of the gradient terms $\sim V^2$ on LHS of full equation (\ref{eq_U}).

Note that the constraint (\ref{long_wl}), which can be considered as the new constraint for the correct use of the previous formulation of TVA, is stronger than the standard constraint
(\ref{TVA_cond_2}) for TVA.

\subsection{The refined general equations for TVA in a disk small shearing sheet\footnote{Note that this Section 
no longer contains an assumption of axial symmetry of the flow.}} 

Section 3 of \citetalias{zhuravlev_2019}, which considers gas-dust dynamics within TVA,
starts from the general equations (18), (19), (13) and (14)\footnote{The corrections below 
should be made to equations (4,9-12) of \citet{zhuravlev_2020} as well.}. It has become clear now, 
that those equations exceed the accuracy of TVA. 
Although \citetalias{zhuravlev_2019} derives the correct final 
equations being completely within TVA, I feel it would be instructive to give here the refined general 
equations for local gas-dust dynamics within TVA, 
as well as the corresponding equations for linear perturbations. 
Since the ordering of terms outlined above has been carried out by \citetalias{zhuravlev_2019} 
with restriction $f\lesssim 1$, it is assumed to be the case below.

\begin{equation}
\label{gen_eq_U_TVA}
\begin{aligned}
(\partial_t - q\Omega_0 x \partial_y) {\bf U} - 2\Omega_0 U_y {\bf e}_x + 
\frac{\kappa^2}{2\Omega_0} U_x {\bf e}_y + ({\bf U}\nabla) {\bf U} = \\ 
\frac{\nabla p_0}{\rho_g} - \frac{\nabla(p+p_0)}{\rho},
\end{aligned}
\end{equation}



\begin{equation}
\label{gen_eq_rho_g_TVA}
\nabla \cdot {\bf U} = 0,
\end{equation}

\begin{equation}
\label{gen_eq_rho_tot_TVA}
(\partial_t - q\Omega_0 x \partial_y) \rho_p + ({\bf U}\nabla) \rho_p + 
t_s \rho\, \nabla\cdot \left ( \rho_p\frac{\nabla(p+p_0)}{\rho^2} \right )  = 0.
\end{equation}

Equations (\ref{gen_eq_U_TVA}-\ref{gen_eq_rho_tot_TVA}) linearised on arbitrary background
should be used instead of equations (A1-A4) of \citetalias{zhuravlev_2019}.

%
%
%
%
%

Taking the smooth background locally representing the solution of \citet{nakagawa-1986},
\begin{equation}
\label{gen_bg_1}
{\bf U} = 0,
\end{equation}
\begin{equation}
\begin{aligned}
\label{gen_bg_2}
\frac{\nabla ( p + p_0 )} {\rho} = {\bf g} \\
\\
\end{aligned}
\end{equation}

and $\rho_p=const$, where the gravity, ${\bf g}$, lies in the $xz$-plane and 
produces the combination of settling and radial drift of the dust,
one derives the corresponding equations for perturbations
\begin{equation}
\label{u}
\begin{aligned}
(\partial_t - q\Omega_0 x \partial_y)\, {\bf u} - 2\Omega_0 u_y {\bf e}_x + 
\frac{\kappa^2}{2\Omega_0} u_x {\bf e}_y = \\ -\nabla W +\frac{f}{1+f} \delta {\bf g},
\end{aligned}
\end{equation}
\begin{equation}
\label{div_u}
\nabla \cdot {\bf u} = 0,
\end{equation}
\begin{equation}
\label{delta}
(\partial_t - q\Omega_0 x \partial_y)\, {\delta} = -t_s \nabla^2 W - \frac{1-f}{1+f} t_s ({\bf g} \nabla) \delta.
\end{equation}

Equations (\ref{u}-\ref{delta}) should be used instead of 
(A5-A7) of \citetalias{zhuravlev_2019}\footnote{Note that the similar corrections should be 
made to equations (B1-B4) and (B8-B11) of \citet{zhuravlev_2020}}.

\section{Conclusion}

The analysis of gas-dust dynamics in the vicinity of resonance between SDW and IW performed using 
the variables $\{{\bf u}_g, W_g, \delta\}$ instead of $\{{\bf u}, W, \delta\}$ clarifies that 
the unstable resonant solution found recently by \citetalias{pan_2021} in the low dust density as well as 
the long wavelength limits arises due to the stationary azimuthal drift of the dust. Since the latter 
physically goes beyond TVA, which was also reflected in the definition of TVA first given by 
\citet{youdin-goodman-2005}, the result of \citetalias{pan_2021}
is explained by an excess of accuracy contained in equations formally known as equations derived using TVA.
The refined general equations describing gas-dust dynamics within TVA are given in Section \ref{sec_gen_eq}.
These equations does not have unstable resonant solutions associated with the low dust density SI.
In turn, this confirms the previous conclusion of \citetalias{zhuravlev_2019} that resonant SI 
goes beyond TVA.
Resonant SI operates due to the combination of two physical effects both associated with the inertia of solids, see the points in the text after equation (\ref{Delta_2}).

\section{Prospects}

The refined equations (\ref{gen_eq_U_TVA}-\ref{gen_eq_rho_tot_TVA}) stand for TVA on the small shearing sheet of a protoplanetary disk. Accordingly, the gas is effectively incompressible.
This model should be generalized on arbitrary dynamics of well-coupled gas-dust 
mixture with the account of compressibility of gas. Particularly, the derivation of 
equations (84-87) of \citet{laibe-price-2014} corresponding to TVA 
should be revisited taking into account the results of this work. 
Most likely, the total density of mixture in the continuity equation (84) of 
\citet{laibe-price-2014} should be replaced by the gas density. Further, it would be interesting to
find the new form of the overstability criterion for partially coupled gas and dust obtained 
by \citet{lin-youdin-2017}, who based their analysis on equations of \citet{laibe-price-2014}
discussed just above. While considering instabilities in dusty protoplanetary disks, 
\citet{lin-youdin-2017} recovered the simplified dispersion equation (27) of \citet{latter-2011},
which partially reproduces the resonant SI as well as the non-resonant growing secular mode. The
present work is focused on the former. However, it is not difficult to check that equations
(22-24, 26) of \citet{latter-2011} along with their equation (25) 
with the last term omitted, which corresponds to the revised TVA, do not reproduce the 
growing secular mode at all, see the analysis in the Appendix \ref{sec_mode} below.
Performing third-order expansion of the full gas-dust equations with respect to the small 
Stokes number,
\citet{jaupart_2020} obtained an accurate growth rate of secular mode and found that 
it reduces to the result of \citet{youdin-goodman-2005} and \citet{latter-2011} in the 
limit $k_x \gg k_z$, see Section 3.2.2 of \citet{jaupart_2020}. Similar reduction with respect 
to the resonant branch of SI as well as the corresponding explanation are shown here, 
see Section \ref{lw_limit} and see also Section 5 of \citetalias{pan_2021}. 
Indeed, the previous formulation of TVA works correctly in the limit of almost radially propagating
harmonics, which corresponds to long waves in the vicinity of resonance, thus, leading to 
agreement with the constraint (\ref{long_wl}). However, the constraint (\ref{long_wl}) 
can be violated well within the known condition of TVA given by equation (\ref{TVA_cond_2}) - 
that is the demonstration of inconsistency of the previous formulation of TVA.
\citet{paardekooper-2020} has used TVA for their study of SI extended onto polydisperse dust. 
They report that wavenumbers of growing modes get larger as compared to SI for the monodisperse dust.
Presumably, this implies that the constraint for the validity of the previous 
formulation of TVA becomes more essential in that case. 
Thus, the constraint (\ref{long_wl}) should be generalized onto polydisperse dust.
On the other hand, the refined equations
for TVA introduced here should be generalized onto the polydisperse dust in order to check whether 
SI is absent within the revised TVA applied to the polydisperse dust.
Finally, it should be noted that the dust-rich branch of SI can be studied employing 
the refined equations for TVA, because at least its linear variant persists in the revised TVA according to the Appendix \ref{dust_rich}.

\section*{Acknowledgments}
The work was supported by the Foundation for the Advancement of Theoretical
Physics and Mathematics ``BASIS'' and the Program of development of Lomonosov Moscow State University.
I acknowledge the support from the Ministry of Science and Higher Education of the Russian Federation (project no. 13.1902.21.0039) for the part of the  study carried out beyond TVA.

\newpage

\bibliography{bibliography}

\appendix

\section{Secular SI mode in the revised TVA}
\label{sec_mode}

Equations (A5-A7) of \citetalias{zhuravlev_2019} yield the dispersion equation
\begin{equation}
\label{disp_app}
\left ( \omega + \frac{g_x k_x t_s}{f+1}  \right ) 
\left ( \omega^2 - \kappa^2 \frac{k_z^2}{k^2} \right ) =
-\frac{g_x k_x t_s}{f+1}\, f \kappa^2 \frac{k_z^2}{k^2} +
\frac{{\rm i}\, f}{f+1}\, \omega^2 t_s \left (\omega^2 - \kappa^2 \right ),
\end{equation}
which is valid for arbitrary value of the dust fraction, $f$.
Equation (\ref{disp_app}) is identical to equation (27) of \citet{latter-2011} taking into 
account that their $f_p = f/(f+1)$ and $f_g = 1/(f+1)$.
Note that the revised TVA provides disappearance of the last term on
RHS of equation (\ref{disp_app}) as can be checked using equations (\ref{u}-\ref{delta}).

Consider the solution of equation (\ref{disp_app}), which satisfies the condition 
$\omega \to 0$ for $t_s \to 0$. As far as $\omega \ll \kappa$, while $k_z \sim k_x$, 
the term $\omega^2$ in the brackets on both sides of equation (\ref{disp_app}) can be omitted. 
To leading order in $t_s$, the corresponding approximate solution reads 
\begin{equation}
\label{sol_app}
\omega \approx - \frac{1-f}{f+1}\, g_x k_x t_s + {\rm i}\, t_s^3 g_x^2\, \frac{f(1-f)^2}{(f+1)^3} 
\frac{k_x^2 k^2}{k_z^2}.
\end{equation}
Equation (\ref{sol_app}) represents exactly the unstable secular mode obtained by \citet{latter-2011}, see their equations (28-29). It is clear from the above derivation 
that the non-zero $\Im[\omega]$ comes from the last term on the RHS of equation (\ref{disp_app}). 
Hence, secular mode obtained within the revised TVA is neutral.

\section{Dust-rich limit of the SI in the revised TVA}
\label{dust_rich}

It can be shown that equation (\ref{disp_app}) has another approximate solution in the high
wavenumber limit, 
\begin{equation}
\label{high_k_app}
g_x k_x t_s \gg \kappa.
\end{equation}
As {\bf long} as $\omega \sim O(\kappa)$, $\omega$ in the first brackets on LHS of equation 
(\ref{disp_app}) can be omitted. Also, the condition (\ref{high_k_app}) implies that 
the second term on the RHS of equation (\ref{disp_app}) becomes small compared to the first term 
over there, as long as $\kappa t_s \lesssim 1$. To zeroth order in small 
$\kappa/(g_x k_x t_s) \ll 1$, this yields the solution 
\begin{equation}
\label{high_f_app}
\omega \approx \pm {\rm i}\, \sqrt{f-1}\, \kappa \frac{k_z}{k}.
\end{equation}
Equation (\ref{high_f_app}) recovers the growing dust-rich mode considered by \citet{squire_2018}, 
see their equation (A2). Hence, the dust-rich limit of SI persists in the more restrictive 
approximation, which is the revised TVA.

\end{document}